\documentclass[twocolumn,pra,aps]{revtex4}
\usepackage{graphicx, color, fancybox}
\usepackage{amsfonts}
\usepackage{amssymb,amsmath}

\begin{document}
\title{Discrimination with error margin between two states \\
- Case of general occurrence probabilities -}
\author{H. Sugimoto, T. Hashimoto, M. Horibe, and A. Hayashi}
\affiliation{Department of Applied Physics\\
           University of Fukui, Fukui 910-8507, Japan}

\begin{abstract}
We investigate a state discrimination problem which interpolates minimum-error 
and unambiguous discrimination by introducing a margin for the probability 
of error. We closely analyze discrimination of two pure states with general occurrence 
probabilities. The optimal measurements are classified into three types. 
One of the three types of measurement is optimal depending on parameters 
(occurrence probabilities and error margin). We determine the three domains  
in the parameter space and the optimal discrimination success probability 
in each domain in a fully analytic form. 
It is also shown that when the states to be discriminated 
are multipartite, the optimal success probability can be attained by 
local operations and classical communication. 
For discrimination of two mixed states,  
an upper bound of the optimal success probability is obtained. 
\end{abstract}

\pacs{PACS:03.67.Hk}
\maketitle

\newcommand{\ket}[1]{|\,{#1}\,\rangle}
\newcommand{\bra}[1]{\langle\,{#1}\,|}
\newcommand{\braket}[2]{\langle\,#1\,|\,#2\,\rangle}
\newcommand{\mbold}[1]{\mbox{\boldmath $#1$}}
\newcommand{\sbold}[1]{\mbox{\boldmath ${\scriptstyle #1}$}}
\newcommand{\tr}[1]{{\rm tr}#1}
\newcommand{\trm}{{\rm tr}}
\renewcommand{\S}{{\rm S}}
\newcommand{\W}{{\rm W}}

\section{Introduction \label{sec:introduction}}
Distinguishing quantum states in various situations is a 
fundamental and highly nontrivial problem in quantum information theory. 
This is because quantum measurement is statistical in nature and it generally destroys 
the state of the system to be measured. 

Quantum state discrimination \cite{Chefles00} is one of such problems. In this problem, 
we are given an unknown quantum state $\rho$, which is chosen from a set of 
known states $\{\rho_a\}$ with some known occurrence probabilities. The task is to 
find the optimal measurement scheme to identify the given state $\rho$ with one in  
the set $\{\rho_a\}$. 
Two settings have been commonly investigated. 
In minimum-error discrimination, the discrimination success probability 
is maximized without any constraint on the probability of erroneous results 
\cite{Helstrom76}. In unambiguous discrimination, however, the success probability 
is maximized under the condition that measurement should not produce erroneous 
results, which is possible by allowing an inconclusive result ``I don't know'' 
\cite{Ivanovic87, Dieks88, Peres88, Jaeger95}. 
Other interesting alternative approaches include a maximum-confidence measurement 
analyzed in Ref.~\cite{Croke06} and the scheme considered in 
Refs.~\cite{Chefles98, Zhang99, Fiurasek03, Eldar_physrev03}, 
in which the probability of correct discrimination is maximized while the rate of 
inconclusive results is fixed.  

We consider a problem of maximizing the success probability under 
the condition that the probability of error should not exceed a certain error margin $m$ 
\cite{Touzel07, Hayashi08}. It is clear that unambiguous discrimination is formulated 
as the case of $m=0$, while minimum-error discrimination corresponds to the case of 
$m=1$. By controlling the error margin, this scheme continuously interpolates 
the minimum-error and unambiguous discrimination problems. 
Touzel, Adamson, and Steinberg \cite{Touzel07} compared 
the numerical results of projective and positive operator-valued measure (POVM) 
measurements in this scheme. In our previous paper \cite{Hayashi08}, 
we analyzed discrimination with error margin between two pure states with equal 
occurrence probabilities and obtained the optimal success probability in a closed 
analytic form. 

In this paper, we extend the analysis of our previous paper \cite{Hayashi08} to the 
case of general occurrence probabilities.  A new feature is that the two-dimensional 
parameter space consisting of occurrence probabilities and the error margin 
is divided into three domains. The types of optimal measurement differ depending 
on the domain. Suppose the error margin is so large that the constraint on 
the probability of error is inactive. Then, the optimal measurement 
is expected to be that of minimum-error discrimination. 
Hereafter, the domain where this is the case is called minimum-error domain. 
To see what happens when the error margin is small, let us recall the results of
unambiguous discrimination ($m=0$). 
If the occurrence probability of one of the states is sufficiently small, 
the optimal measurement produces only two outcomes omitting this state. 
For general error margin, this is expected to happen in a domain of the parameter 
space, which we call single-state domain. Intermediate domain is the one where 
probabilities of three measurement outcomes are non zero.

The main purpose of this paper is to determine these three domains and the optimal 
success probability in each domain in a fully analytic form. The problem is formulated and 
the main results are presented in Sec.~\ref{sec:problem}. Derivation of the results is detailed 
in Secs.~\ref{sec:intermediate} and \ref{sec:single}. 

We can consider two types of error margin for the probability of error. 
One is the constraint on the mean probability of error, which will be discussed first. 
The other is the constraint on conditional error probabilities. 
In Sec.~\ref{sec:strong}, we establish a relation between the optimal success probabilities 
of the two types of constraint. 
We also discuss discrimination of two mixed states. 
In Sec.~\ref{sec:mixed}, we show that an upper bound of the success probability 
for two mixed states can easily be obtained in terms of the optimal success probability 
of two pure states. 
\section{Problem and solution \label{sec:problem}} 
We consider the discrimination problem between two pure states 
$\rho_1=\ket{\phi_1}\bra{\phi_1}$ and $\rho_2=\ket{\phi_2}\bra{\phi_2}$ 
with occurrence probabilities $\eta_1$ and $\eta_2$, respectively.  
To avoid trivial exceptional cases, we assume that $\eta_1 \ne 0$ and $\eta_2 \ne 0$. 
We also assume that the two states are linearly independent and we work in the 
two-dimensional subspace $V$ spanned by these two states. The measurement 
is described by a positive operator-valued measure on $V$, 
which consists of three elements $\{E_\mu\}_{\mu=1}^3$. 
Measurement outcome labeled by $\mu=1$ or $2$ means that the given input state 
is identified with state $\rho_\mu$. Element $E_3$ produces the inconclusive result. 
Let us denote by $P_{\rho_a,E_\mu}$ the joint probability 
that the given state is $\rho_a\ (a=1,2)$ and the measurement outcome is $\mu$. 
The probability $P_{\rho_a,E_\mu}$ is given by 
\begin{equation*}
  P_{\rho_a,E_\mu} = \eta_a \tr{E_\mu\rho_a}. 
\end{equation*}

The discrimination success probability $p_{\circ}$ and the mean probability of error 
$p_{\times}$ are given by 
\begin{align}
  p_{\circ} & \equiv  P_{\rho_1,E_1} + P_{\rho_2,E_2} \nonumber \\
            & =       \eta_1 \tr{E_1 \rho_1} + \eta_2 \tr{E_2 \rho_2}, \\
  p_{\times}& \equiv  P_{\rho_1,E_2} + P_{\rho_2,E_1} \nonumber \\
            & =       \eta_1 \tr{E_2 \rho_1} + \eta_2 \tr{E_1 \rho_2}. 
\end{align} 
We require that the mean probability of error $p_{\times}$ must not exceed a 
certain error margin $m\ (0 \le m \le 1)$.  Then, the task is to maximize the success 
probability $p_{\circ}$ under the conditions: 
\begin{subequations} 
  \label{eq:primal}
\begin{align}
& E_1 \ge 0,\ \ E_2 \ge 0,\ \ E_3 \ge 0, \label{eq:primal_POVM_positivity}\\ 
& E_1+E_2+E_3 = 1, \label{eq:primal_POVM_completeness} \\
& p_{\times} \le m, \label{eq:primal_weak_margin}
\end{align}
\end{subequations}
where Eqs.~(\ref{eq:primal_POVM_positivity}, \ref{eq:primal_POVM_completeness}) are the 
usual conditions for a POVM. 

This problem can be formulated as one of semidefinite programming (SDP). 
See Ref.~\cite{Vandenberghe96} for a general review and 
Refs.~\cite{Eldar_Megretski03, Eldar_IEEE03} for applications of SDP to quantum-state 
discrimination. According to the general theory of SDP, we can write the necessary and 
sufficient conditions for the optimal POVM. For our purpose, it suffices to see 
that they are sufficient conditions.  

Suppose a Hermitian operator $Y$ acting on $V$ and a real number $y$ satisfy conditions 
\begin{subequations}
    \label{eq:upper}
\begin{align}
  Y & \ge  0, \label{eq:upper_Y} \\
  Y & \ge  \eta_1\rho_1-y\eta_2\rho_2, \label{eq:upper_Y_1} \\
  Y & \ge  \eta_2\rho_2-y\eta_1\rho_1, \label{eq:upper_Y_2} \\
  y & \ge  0. \label{eq:upper_y}
\end{align}
\end{subequations}
It is easy to show that 
\begin{equation}
   d \equiv \tr{Y}+my, 
\end{equation}
gives an upper bound for the success probability $p_{\circ}$, because 
\begin{align*}
  p_{\circ} &= \eta_1 \tr{E_1 \rho_1} + \eta_2 \tr{E_2 \rho_2}  \\
            & \le  \tr{E_1(Y+y\eta_2\rho_2)} + \tr{E_2(Y+y\eta_1\rho_1)} \\ 
            & =  \tr{(E_1+E_2)Y} + yp_{\times} \\
            & \le  \tr{Y}+ym = d.
\end{align*}
It is clear that this upper bound is attained if and only if the following 
relations hold: 
\begin{subequations}
  \label{eq:attain}
\begin{align}
  & E_1(Y-(\eta_1\rho_1-y\eta_2\rho_2)) = 0, \label{eq:attain_E_1}\\ 
  & E_2(Y-(\eta_2\rho_2-y\eta_1\rho_1)) = 0, \label{eq:attain_E_2}\\
  & E_3Y = 0, \label{eq:attain_E_3} \\
  & y(m-p_{\times}) =0. \label{eq:attain_y}
\end{align}
\end{subequations}

Thus, the set of equations given by Eqs.~(\ref{eq:primal}), (\ref{eq:upper}), 
and (\ref{eq:attain}) is a sufficient condition for an optimal solution. 
As we will see, we can construct a solution satisfying this condition for 
any parameters: $\eta_a,m$ and $\braket{\phi_1}{\phi_2}$.   
The general theory SDP shows it is also a necessary condition 
\cite{Vandenberghe96,Eldar_Megretski03, Eldar_IEEE03}.   
Minimizing $d$ under conditions Eqs.~(\ref{eq:upper}) is called dual problem, 
whereas the original problem of maximizing $p_{\circ}$ under conditions 
Eqs.~(\ref{eq:primal}) is referred to as primal problem. 

Let us begin by looking at ranks of optimal POVM elements, which are operators 
on the two-dimensional space $V$. 
We note that they are of rank 1 at most. 
This can be seen in the following way. Suppose that $E_3$ is of rank 2. Condition 
Eq.~(\ref{eq:attain_E_3}) requires that $Y=0$. 
Then, from Eqs.~(\ref{eq:upper_Y_1}, \ref{eq:upper_Y_2}), we find 
$y\eta_2\rho_2 \ge \eta_1\rho_1$ and $y\eta_1\rho_1 \ge \eta_2\rho_2$. 
It is easy to see that these inequalities contradict the assumption that the two states 
are linearly independent. 
Next, suppose that $E_1$ is of rank 2.  From Eq.~(\ref{eq:attain_E_1}), we have 
$Y=\eta_1\rho_1-y\eta_2\rho_2$. 
Then, Eqs.~(\ref{eq:upper_Y}, \ref{eq:upper_Y_2}) require that 
$\eta_1\rho_1 \ge y\eta_2\rho_2$ and $\eta_1\rho_1 \ge \eta_2\rho_2$, 
which are again inconsistent with the linear independence of the two states and 
the assumption that $\eta_2 \ne 0$. It is clear that the rank of $E_2$ is also 
1 at most. 

As stated in Sec.~\ref{sec:introduction}, there are three types of measurements, 
one of which becomes optimal depending on domains of the parameter space of occurrence 
probabilities and error margin. This classification can be done according to 
the ranks of POVM.  In the minimum-error domain, the optimal POVM is that of 
minimum-error discrimination, which implies that ranks of $E_1$ and $E_2$ is 1 
while $E_3=0$. In the single-state domain, optimal measurement produces only 
two outcomes omitting one of the two states. In this case, either $E_1$ or $E_2$ is 0 
and the remaining two POVM elements are of rank 1. The intermediate domain is where 
all POVM elements are of rank 1 and probabilities of obtaining the three outcomes 
are non zero. 

In what follows, we present the main results first, leaving their derivation to subsequent  
sections. We assume that $\eta_1 \le \eta_2$ without loss of generality. 
To make expressions simpler, we define 
\begin{align}
  S & \equiv  |\braket{\phi_1}{\phi_2}|^2, 
            \label{eq:S} \\
  T & \equiv  1- |\braket{\phi_1}{\phi_2}|^2. 
            \label{eq:T} 
\end{align} 

\begin{figure}
\includegraphics[width=0.8\hsize]{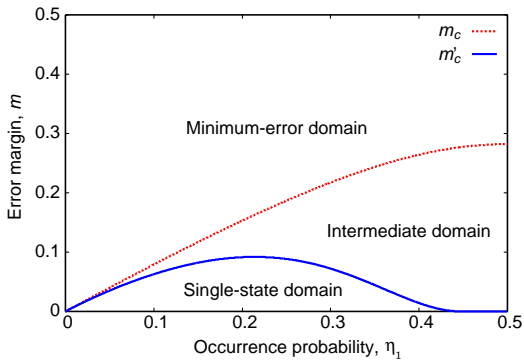}
\caption{\label{fig:domain}
The three domains in the parameter space of occurrence probability $\eta_1$ and error margin $m$: 
minimum-error domain, intermediate domain, and single-state domain.
Fidelity $|\braket{\phi_1}{\phi_2}|$ is taken to be $0.9$. 
}
\end{figure}

\begin{figure}
\includegraphics[width=0.8\hsize]{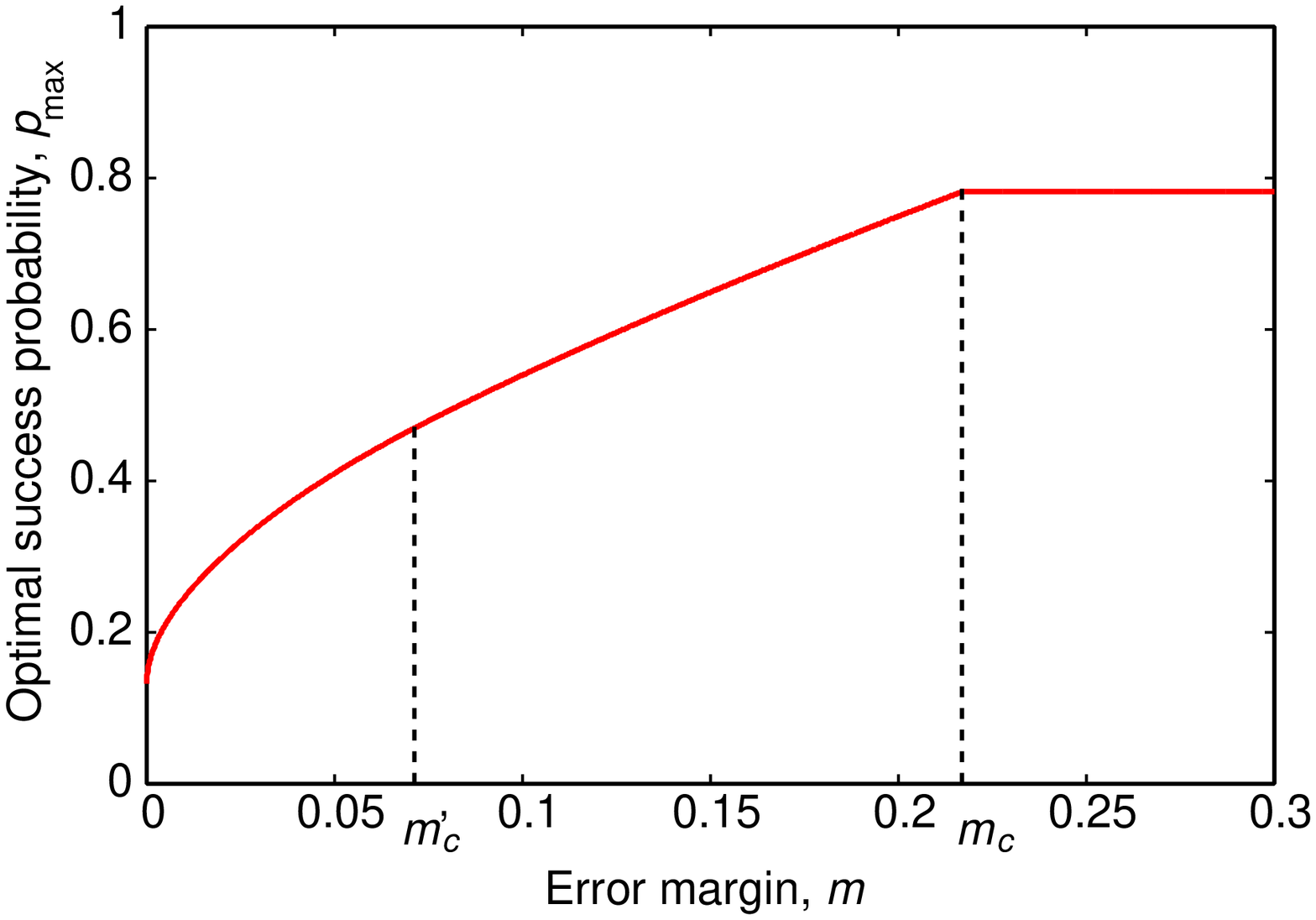} \\
\includegraphics[width=0.8\hsize]{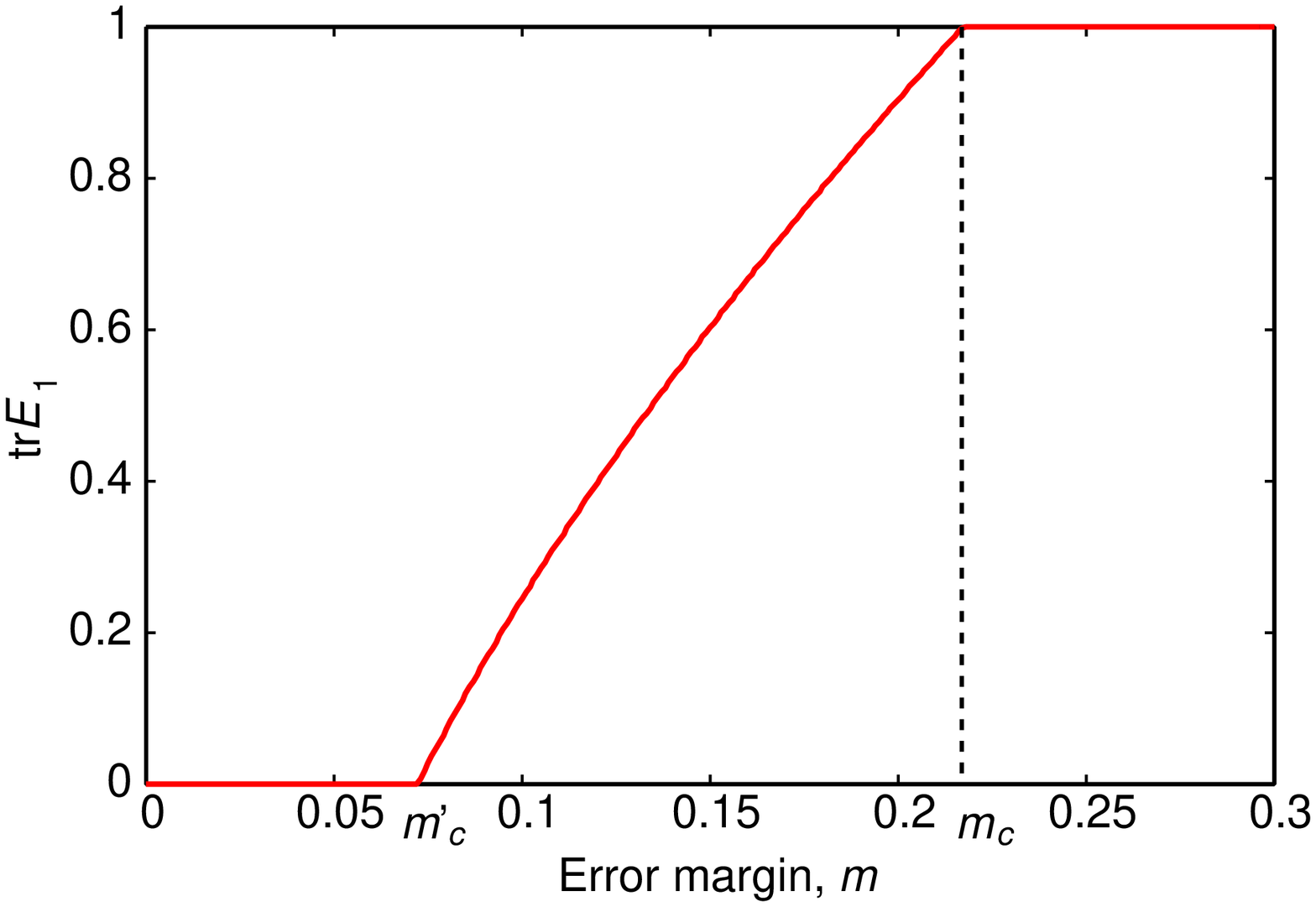} 
\caption{\label{fig:curve}
The optimal success probability $p_{\max}$ (upper part) and $\tr E_1$ (lower part) 
vs error margin $m$.
The occurrence probability $\eta_1$ is 0.3 and 
fidelity $|\braket{\phi_1}{\phi_2}|$ is $0.9$. 
}
\end{figure}

The parameter space is divided into the following three domains: 
\begin{equation*}
    \begin{cases} 
      \text{Minimum-error domain : } & m_c \le m \le 1, \\
      \text{Intermediate domain : }  & m_c' \le m \le m_c,  \\
      \text{Single-state domain : }  & 0 \le m \le m_c', 
    \end{cases}
\end{equation*}
where two critical error margins $m_c$ and $m_c'$ are defined by 
\begin{align}
    m_c & \equiv  \frac{1}{2}\left( 1 - \sqrt{1-4\eta_1\eta_2 S} \right),  
                   \label{eq:mc} \\
    m_c'& \equiv  
            \begin{cases} 
               \displaystyle
               \frac{(\eta_1-\sqrt{\eta_1\eta_2S})^2}{1-2\sqrt{\eta_1\eta_2S}} &
                            (\eta_1 \le \eta_2 S), \\
               0 & (\eta_1 \ge \eta_2 S).
            \end{cases}
                   \label{eq:mc_prime}
\end{align} 
Figure \ref{fig:domain} depicts the three domains in the case of 
$\sqrt{S}=|\braket{\phi_1}{\phi_2}|=0.9$.  
The optimal discrimination success probability in each domain is found to be 
\begin{equation}
  p_{\max} =  
     \begin{cases} 
        \frac{1}{2}\left( 1 + \sqrt{1-4\eta_1\eta_2 S} \right) &
                            (m_c \le m \le 1), \\ 
        \left( \sqrt{m} + \sqrt{ 1 - 2\sqrt{\eta_1\eta_2 S} } \right)^2  &
                            (m_c' \le m \le m_c), \\ 
        \eta_2 \left( \sqrt{\frac{m}{\eta_1}S} + \sqrt{\frac{\eta_1-m}{\eta_1}T} \right)^2 &  
                            (0 \le m \le m_c'). 
     \end{cases}
           \label{eq:weak_maximum_p_circ}
\end{equation}
The critical margin $m_c$ is actually the mean error probability of optimal  
minimum-error discrimination. If $m_c \le m$, the constraint on the probability of error  
is inactive.  This is the reason why $p_{\max}$ in the minimum-error domain is given by 
that of minimum-error discrimination. 
In Fig.~\ref{fig:curve}, we plot the optimal success probability $p_{\max}$ and 
$\tr E_1$ against error margin $m$ for a fixed $\eta_1$. 
The plot of $\tr E_1$ clearly shows the border between the single-state and 
intermediate domains, though the curve of $p_{\max}$ is smooth at $m=m_c'$. 

In unambiguous discrimination ($m=0$), for a sufficiently small $\eta_1$, 
the optimal measurement is always of the single-state type. Intuitively, this appears  
reasonable. However, Fig.~\ref{fig:domain} shows that this is no longer true for a 
finite error margin. For example, fix $m$ to be around 0.06 and vary $\eta_1$ 
from 0.5 to 0. Then, the type of optimal measurement varies in a nontrivial 
way: from the intermediate to single-state, intermediate, and minimum-error type. 

\begin{figure}
\includegraphics[width=0.8\hsize]{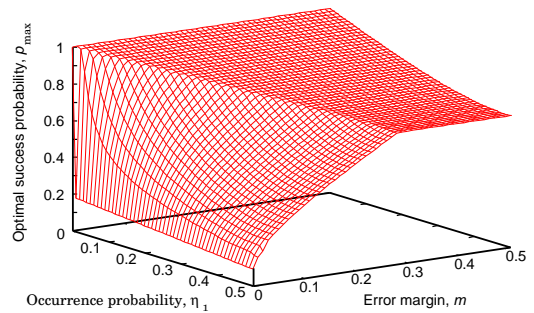}
\caption{\label{fig:3D}
Three-dimensional plot of the optimal success probability $p_{\max}$ 
vs occurrence probability $\eta_1$ and error margin $m$. 
Fidelity $|\braket{\phi_1}{\phi_2}|$ is taken to be $0.9$. 
}
\end{figure}

Figure \ref{fig:3D} displays a three-dimensional overview of the optimal  
success probability. 

\section{Intermediate domain \label{sec:intermediate}}
In this section, we construct a solution where all POVM elements $E_1$, $E_2$, and $E_3$ are 
non zero and of rank 1. The attainability conditions given by 
Eqs.~(\ref{eq:attain_E_1}-\ref{eq:attain_E_3}) imply that 
positive semidefinite operators 
\begin{align*}
  Y_1 & \equiv  Y-(\eta_1\rho_1-y\eta_2\rho_2),  \\
  Y_2 & \equiv  Y-(\eta_2\rho_2-y\eta_1\rho_1), 
\end{align*} 
and $Y$ are all of rank 1. 
It is convenient to use the Bloch vector representation for 
$\rho_a$ and other operators acting on $V$. 
\begin{equation*}
   \rho_a = \frac{1+\mbold{n}_a \cdot \mbold{\sigma}}{2}\ (a=1,2),
\end{equation*}
where $\mbold{\sigma}=(\sigma_x,\sigma_y,\sigma_z)$ are Pauli's matrices.    
Writing  
\begin{equation*}
  Y = \alpha + \mbold{\beta} \cdot \mbold{\sigma}, 
\end{equation*}
we have
\begin{align*}
  Y_1 &= \alpha-\frac{\eta_1-y\eta_2}{2} + 
          \left( \mbold{\beta} - \frac{\mbold{a}_1}{2} \right) \cdot \mbold{\sigma}, \\
  Y_2 &= \alpha-\frac{\eta_2-y\eta_1}{2} + 
          \left( \mbold{\beta} - \frac{\mbold{a}_2}{2} \right) \cdot \mbold{\sigma},          
\end{align*}
where we introduced two vectors $\mbold{a}_1$ and $\mbold{a}_2$ defined to be  
\begin{align}
  \mbold{a}_1 &= \eta_1 \mbold{n}_1-y \eta_2 \mbold{n}_2, \label{eq:a_1}\\
  \mbold{a}_2 &= \eta_2 \mbold{n}_2-y \eta_1 \mbold{n}_1. \label{eq:a_2}
\end{align}
Since the smaller eigenvalues of operators $Y_1$, $Y_2$, and $Y$ are all zero, 
we obtain the following three equations for $\alpha$ and $\mbold{\beta}$:  
\begin{subequations}
   \label{eq:beta_equation}
\begin{align}
 & \alpha - \frac{\eta_1-y\eta_2}{2} 
      = \left| \mbold{\beta} - \frac{\mbold{a}_1}{2} \right|, \\
 & \alpha - \frac{\eta_2-y\eta_1}{2} 
      = \left| \mbold{\beta} - \frac{\mbold{a}_2}{2} \right|, \\
 & \alpha = |\mbold{\beta}|. 
\end{align} 
\end{subequations}

Solving Eqs.~(\ref{eq:beta_equation}) requires a rather long calculation. 
It turns out that parameter $y$ must satisfy $y \ge 1$ and  
vector $\mbold{\beta}$ is given by 
\begin{align}
   \mbold{\beta}  = \frac{y}{2(y-1)} 
      \bigg\{ 
          & \left( \eta_1 \pm \sqrt{\frac{\eta_1\eta_2}{S}} \right) \mbold{n}_1 
                              \nonumber \\  
          & + \left( \eta_2 \pm \sqrt{\frac{\eta_1\eta_2}{S}} \right) \mbold{n}_2 
      \bigg\},
        \label{eq:vector_beta}
\end{align}
and $\alpha$ is given by
\begin{equation}
   \alpha = |\mbold{\beta}| =\frac{y}{2(y-1)} \left( 1 \pm 2\sqrt{\eta_1\eta_2S} \right).  
        \label{eq:alpha}
\end{equation}

Now that we have $Y$ and $y$ satisfying Eqs.~(\ref{eq:upper}), we obtain  
an upper bound for the success probability by calculating $d=\tr{Y}+ym$. 
\begin{equation}
  d = \tr{Y}+ym = \frac{y}{y-1} \left( 1 \pm 2\sqrt{\eta_1\eta_2 S} \right) + my.
            \label{eq:intermediate_upper_before_minimization}
\end{equation}
We determine parameter $y$ so that the upper bound $d$ is minimized, which leads to 
\begin{align}
    d &= \left( \sqrt{m} + \sqrt{ 1 \pm 2\sqrt{\eta_1\eta_2 S} } \right)^2,  
                  \label{eq:intermediate_upper} \\
    y &= 1 + \frac{\sqrt{ 1 \pm 2\sqrt{\eta_1\eta_2 S}}}{\sqrt{m}}.
\end{align}
As to the double signs in the above equations, we take a negative one to obtain a 
smaller upper bound. Correspondingly, a negative sign is taken also in double signs of  
Eqs.~(\ref{eq:vector_beta},\ref{eq:alpha}) hereafter.  

The attainability conditions given by 
Eqs.~(\ref{eq:attain_E_1}-\ref{eq:attain_E_3}) require that $E_1$, $E_2$, and $E_3$ 
take the following form:
\begin{equation*}
  E_{\mu} = \gamma_\mu \left( |\mbold{\beta}_\mu|-\mbold{\beta}_\mu \cdot \sigma \right), 
                         \ (\mu=1,2,3),
                         \label{eq:POVM_candidate}
\end{equation*}
where we defined 
\begin{equation*}
  \mbold{\beta}_1 \equiv \mbold{\beta}-\frac{1}{2}\mbold{a}_1,\ 
  \mbold{\beta}_2 \equiv \mbold{\beta}-\frac{1}{2}\mbold{a}_2,\ 
  \mbold{\beta}_3 \equiv \mbold{\beta}. 
\end{equation*}  
The question is whether positive constants $\gamma_1$, $\gamma_2$, and $\gamma_3$ can be 
chosen so that the set $\{E_1,E_2,E_3\}$ respects the completeness condition of POVM 
given in Eq.~(\ref{eq:primal_POVM_completeness}). This is possible if and only if 
a linear relation with positive coefficients exists for 
three vectors $\mbold{\beta}_1$, $\mbold{\beta}_2$, and $\mbold{\beta}_3$. 
\begin{equation*}
   c_1 \mbold{\beta}_1 + c_2 \mbold{\beta}_2 +c_3 \mbold{\beta}_3  = 0,\ 
  (c_1,c_2,c_3 \ge 0). 
\end{equation*}
If such a linear relation exists, coefficients $\gamma_\mu$ can be constructed as 
$ \gamma_\mu = \gamma c_\mu$ with an overall positive factor $\gamma$ determined 
so that $\sum_{\mu} \gamma_\mu |\mbold{\beta}_\mu|=1$. 

Since each of the three vectors is expressed by the two Bloch vectors $\mbold{n}_1$ and 
$\mbold{n}_2$,  a linear relation, which is unique up to an overall factor, is 
straightforwardly found, with coefficients given by 
\begin{align*}
  c_1 &= \frac{y}{y+1}
          \left( \sqrt{m} - \frac{\sqrt{\eta_1\eta_2S}-\eta_1}
                                  {\sqrt{1-2\sqrt{\eta_1\eta_2S}}} \right), \\
  c_2 &= \frac{y}{y+1}
          \left( \sqrt{m} - \frac{\sqrt{\eta_1\eta_2S}-\eta_2}
                                  {\sqrt{1-2\sqrt{\eta_1\eta_2S}}} \right), \\
  c_3 &= \sqrt{\frac{\eta_1\eta_2S}{m}} 
          - \sqrt{m} - \sqrt{1-2\sqrt{\eta_1\eta_2S}} .                                   
\end{align*}
Signs of $c_\mu$ vary depending on $\eta_1$, $\eta_2$, $S$, and $m$. 
Remember that we assumed $\eta_1 \le \eta_2$. Then $c_2$ is always positive. 
We find that $c_1$ is positive if $m \ge m_c'$ and $c_3$ is positive if 
$m \le m_c$, with $m_c$ and $m_c'$ defined in Eqs.~(\ref{eq:mc}) and (\ref{eq:mc_prime}), 
respectively. Thus, the set $\{E_1,E_2,E_3\}$ is a POVM if the error margin 
is in the range $m_c' \le m \le m_c$. 

Remaining conditions are Eq.~(\ref{eq:primal_weak_margin}) and Eq.~(\ref{eq:attain_y}), 
which are reduced to $p_{\times}=m$ since $y \ge 1$. 
We can explicitly verify that the relation $p_{\times}=m$ holds after a long calculation 
by using the POVM constructed above. 
This is not a coincidence, but a consequence of how we determined parameter $y$. 
Parameter $y$ was determined so that the upper bound $d$ given by 
Eq.~(\ref{eq:intermediate_upper_before_minimization}) is minimized:
\begin{equation*}
   \frac{\partial}{\partial y}d = \frac{\partial}{\partial y} \tr{Y} +m =0. 
\end{equation*}
We can show that $\frac{\partial}{\partial y} \tr{Y} = -p_{\times}$, which 
means that minimization of $d$ leads to the relation $p_{\times}=m$. 
This can be seen in the following way. 
Suppose two positive semidefinite operators $A(y)$ and $B(y)$ depend on a 
variable $y$ and satisfy $A(y)B(y)=0$. 
Then we can show $\tr A(y) B^{'}(y) = 0$.  
To prove this, we define a function $f(x)$ to be 
\begin{equation*}
  f(x) \equiv \tr A(y) B(y+x).  
\end{equation*}
Note that $f(0) =0$ while $f(x) \ge 0$ for any $x$, which implies that $f(x)$ has a 
minimum at $x=0$. From $f^{'}(0)=0$, the desired result immediately follows.   
Now, operators $Y_1$, $Y_2$, $Y$ and POVM elements $E_\mu$ are all positive 
semidefinite and satisfy $E_1Y_1=E_2Y_2=E_3Y=0$. We therefore obtain 
\begin{align*}
 \frac{\partial}{\partial y} \tr Y 
 &=   \tr E_1 \frac{\partial Y}{\partial y} 
      +\tr E_2 \frac{\partial Y}{\partial y} 
      +\tr{E_3 \frac{\partial Y}{\partial y}} \\
 &=   \tr E_1 \frac{\partial Y_1}{\partial y}  
      +\tr E_2 \frac{\partial Y_2}{\partial y}  
      +\tr E_3 \frac{\partial Y}{\partial y} - p_{\times} \\
 &= -p_{\times}.
\end{align*}

Thus, if error margin $m$ is in the range $m_c' \le m \le m_c$, the upper bound 
of Eq.~(\ref{eq:intermediate_upper}) is attained and the maximum success probability  
is given by 
\begin{equation}
  p_{\max} = \left( \sqrt{m} + \sqrt{ 1 - 2\sqrt{\eta_1\eta_2 S} } \right)^2. 
             \label{eq:intermediate_p_max}
\end{equation}

By using the optimal POVM, we find that the following symmetries turn out to hold:
\begin{align}
  P_{\rho_2 | E_1} &= P_{\rho_1 | E_2}, \label{eq:conditional_error_symmetry} \\ 
  P_{\rho_1 | E_3} &= P_{\rho_2 | E_3}, \label{eq:E_3_symmetry} 
\end{align}
where we introduced conditional probabilities defined by   
\begin{equation*}
 P_{\rho_a|E_\mu} \equiv \frac{P_{\rho_a,E_\mu}}{P_{E_\mu}},\ 
 P_{E_\mu} \equiv P_{\rho_1,E_\mu}+P_{\rho_2,E_\mu}. 
\end{equation*} 
This is noteworthy, since in the problem, there is no apparent symmetry between 
$\rho_1$ and $\rho_2$ with general occurrence probabilities. 
The symmetry between two conditional error probabilities given by Eq.~(\ref{eq:conditional_error_symmetry}) will be important in Sec.~\ref{sec:strong}. 

Before concluding the section, we present a simple argument to clarify  
how these symmetries emerge. Let us define two vectors $\mbold{C}$ and 
$\mbold{X}$ to be  
\begin{align*}
  \mbold{C} & \equiv  \left( \sqrt{P_{\rho_1,E_1}}, \sqrt{P_{\rho_2,E_2}} \right), \\
  \mbold{X} & \equiv  \left( \sqrt{P_{\rho_2,E_1}}, \sqrt{P_{\rho_1,E_2}} \right).
\end{align*}
The success probability $p_{\circ}$ is then given by $|\mbold{C}|^2$.  
By the triangle inequality we observe 
\begin{equation}
   \sqrt{p_{\circ}} = |\mbold{C}| \le |\mbold{X}|+|\mbold{C-X}|. 
          \label{eq:triangle_inequality} 
\end{equation}
Note $|\mbold{X}|=\sqrt{p_{\times}}$, which must not exceed $\sqrt{m}$. 
An upper bound of $|\mbold{C-X}|$ can be determined in the following way.  
\begin{align*}
  |\mbold{C-X}|^2 &= \mbold{C}^2+\mbold{X}^2-2\mbold{C}\cdot\mbold{X} \\
   &= 1 - P_{\rho_1,E_3} - P_{\rho_2,E_3} - 2\mbold{C}\cdot\mbold{X} \\
   & \le  
      1-2\left(  \sqrt{P_{\rho_1,E_3}P_{\rho_2,E_3}} + \mbold{C}\cdot\mbold{X} \right) \\
   &= 1-2\sqrt{\eta_1\eta_2} \sum_{\mu=1}^3 \sqrt{q_{\mu}^{(1)}q_{\mu}^{(2)}},
\end{align*}  
where we used the inequality of arithmetic and geometric means  
\begin{equation}
  P_{\rho_1,E_3} + P_{\rho_2,E_3} \ge 2\sqrt{P_{\rho_1,E_3}P_{\rho_2,E_3}}, 
              \label{eq:arithmetic_mean}
\end{equation}
and we defined two probability distributions $q_{\mu}^{(1)}$ and $q_{\mu}^{(2)}$ by 
\begin{equation*}
  q_{\mu}^{(a)} \equiv \tr \rho_a E_{\mu}\ (a=1,2,\ \mu=1,2,3).  
\end{equation*}
Expression $\sum_{\mu=1}^3 \sqrt{q_{\mu}^{(1)}q_{\mu}^{(2)}}$ is the fidelity 
of two classical probability distributions $q_{\mu}^{(1)}$ and $q_{\mu}^{(2)}$ 
of obtaining measurement outcome $\mu$ for the two state $\rho_1$ and $\rho_2$.
This classical fidelity is known to be lower-bounded by the quantum fidelity 
of the two states  
$|\braket{\phi_1}{\phi_2}|=\sqrt{S}$ \cite{Nielsen_text_book}. 
Thus, we obtain an upper bound for $p_\circ$ as 
\begin{equation*}
  p_\circ \le \left( \sqrt{m} + \sqrt{ 1 - 2\sqrt{\eta_1\eta_2 S} } \right)^2. 
\end{equation*}
We notice that this is the attainable maximum given by Eq.~(\ref{eq:intermediate_p_max}). 
Consequently, equality must holds in all inequalities used to obtain this upper bound. 
Among them, equality of the triangle inequality in Eq.~(\ref{eq:triangle_inequality}) implies  
vectors $\mbold{C}$ and $\mbold{X}$ are in the same direction, which immediately leads to 
the symmetry of Eq.~(\ref{eq:conditional_error_symmetry}).  Equality of inequality 
(\ref{eq:arithmetic_mean}) requires the relation of Eq.~(\ref{eq:E_3_symmetry}). 

\section{Single-state domain \label{sec:single}}
In unambiguous discrimination ($m=0$), omitting one of the states to be 
discriminated is optimal if its occurrence probability is sufficiently small. 
In discrimination with general error margin, a similar situation occurs in a domain of 
parameters ($\eta_1$ and $m$), which we call single-state domain.  
In this section, we will determine the optimal success probability in the single-state domain.

Assuming $\eta_1 \le \eta_2$, we search for optimal POVM with $E_1=0$. 
Remember that all POVM elements are of rank 1 at most. We immediately see 
that $E_2$ and $E_3$ must constitute a projective measurement 
with respect to a set of orthonormal states $\ket{\mbold{f}}$ and $\ket{-\mbold{f}}$, 
with $\mbold{f}$ being a unit Bloch vector to be determined. 
\begin{align}
  E_2 &= \ket{\mbold{f}}\bra{\mbold{f}}, \\
  E_3 &= \ket{-\mbold{f}}\bra{-\mbold{f}}.
\end{align}
Now look at the attainability conditions Eq.~(\ref{eq:attain}). Equation (\ref{eq:attain_E_1}) 
is trivially satisfied. Equations (\ref{eq:attain_E_2}) and (\ref{eq:attain_E_3}) require that 
\begin{align}
  & Y = \lambda_{+} \ket{\mbold{f}}\bra{\mbold{f}}, \label{eq:Y_lambda} \\
  & Y-(\eta_2\rho_2-y\eta_1\rho_1) = -\lambda_{-}\ket{-\mbold{f}}\bra{-\mbold{f}}, 
                                                      \label{eq:Y_2_lambda} 
\end{align}
where $\lambda_{+}$ and $\lambda_{-}$ are constants. We see that 
$\lambda_{+} \ge 0$ and $\lambda_{-} \le 0$ from upper bound conditions 
Eq.~(\ref{eq:upper_Y}) and Eq.~(\ref{eq:upper_Y_2}). 
Eliminating $Y$ from Eqs.~(\ref{eq:Y_lambda}) and (\ref{eq:Y_2_lambda}), we find 
\begin{equation*}
  \eta_2\rho_2-y\eta_1\rho_1 = \lambda_{+}\ket{\mbold{f}}\bra{\mbold{f}}
                              +\lambda_{-}\ket{-\mbold{f}}\bra{-\mbold{f}}, 
\end{equation*}
which is the spectral decomposition of operator $\eta_2\rho_2-y\eta_1\rho_1$. 
This shows that $\lambda_{+}$ and $\lambda_{-}$ are the positive and negative eigenvalues with 
eigenstates $\ket{\mbold{f}}$ and $\ket{-\mbold{f}}$, respectively.   
We thus obtain $\lambda_{+}$, $\lambda_{-}$, and $\mbold{f}$ in terms of 
Bloch vectors $\mbold{n}_1$ and $\mbold{n}_2$. 
\begin{align}
  \lambda_{\pm} &= \frac{1}{2}(\eta_2-y\eta_1) \pm \frac{1}{2} |\mbold{a}_2|, \\
  \mbold{f} &= \frac{\mbold{a}_2}{|\mbold{a}_2|},
\end{align}
where $\mbold{a}_2 = \eta_2\mbold{n}_2-y\eta_1\mbold{n}_1$ as defined in Eq.~(\ref{eq:a_2}).       

Parameter $y$ still remains to be determined. This can be done by requiring conditions 
Eqs.~(\ref{eq:primal_weak_margin}), ({\ref{eq:upper_Y_1}}), ({\ref{eq:upper_y}}), 
and ({\ref{eq:attain_y}}), which have not been checked so far. 

The positivity of $Y-(\eta_1\rho_1-y\eta_2\rho_2)$ of Eq.~(\ref{eq:upper_Y_1}) can be  
expressed as 
\begin{equation*}
 \lambda_{+}-(\eta_1-y\eta_2) \ge 
    \left| \lambda_{+}\frac{\mbold{a}_2}{|\mbold{a}_2|}-\mbold{a}_1 \right|,
\end{equation*}    
where $\mbold{a}_1 = \eta_1\mbold{n}_1-y\eta_2\mbold{n}_2$ and        
$\mbold{a}_2 = \eta_2\mbold{n}_2-y\eta_1\mbold{n}_1$.       
After a rather involved calculation, we find that 
this condition together with positivity of $y$, Eq.~(\ref{eq:upper_y}),  
imply the occurrence probabilities must satisfy an inequality given by 
\begin{equation}
  \eta_1 \le \eta_2 S, 
             \label{eq:single_eta_range}
\end{equation}
and parameter $y$ an inequality given by 
\begin{equation}
  y \ge 1+\frac{(1-2\sqrt{\eta_1\eta_2 S})(1+\sqrt{\frac{\eta_2 S}{\eta_1}})}{\eta_2 S-\eta_1}.
             \label{eq:single_y_range}  
\end{equation}
The remaining conditions Eqs.~(\ref{eq:primal_weak_margin}) and (\ref{eq:attain_y}) are 
simply reduced to a single equation $p_{\times}=m$ since $y \ge 1$ by Eq.~(\ref{eq:single_y_range}).  
The average probability of error $p_{\times}$ is calculated as 
\begin{equation*}
  p_{\times} = \eta_1 \tr{E_2 \rho_1} = \eta_1\frac{1+\mbold{f}\cdot \mbold{n}_1}{2}, 
\end{equation*}
which should be equated to error margin $m$. This establishes a relation between parameter $y$ 
and error margin $m$.
\begin{equation}
  y = \frac{\eta_2}{\eta_1} \left( 
          T-S + \sqrt{ST} \frac{\eta_1-2m}{\sqrt{m(\eta_1-m)}} \right).
                \label{eq:single_ym_relation}
\end{equation}  
We can now translate the allowed range of parameter $y$ given in 
Eq.~(\ref{eq:single_y_range}) to that of error margin $m$. We find that 
the allowed range of error margin is given by 
\begin{equation}
    0 \le m \le \frac{(\eta_1-\sqrt{\eta_1\eta_2S})^2}{1-2\sqrt{\eta_1\eta_2S}}. 
         \label{eq:single_m_range}
\end{equation}
Combining this with the condition (\ref{eq:single_eta_range}), we see that  
the single-state domain is specified by inequality $0 \le m \le m_c'$,
with $m_c'$ defined in Eq.~(\ref{eq:mc_prime}).

The optimal success probability in the single-state domain is obtained by calculating 
$d=\tr{Y}+ym$.  
\begin{equation}
  p_{\max} = \eta_2 \left( 
        \sqrt{\frac{m}{\eta_1}S} + \sqrt{\frac{\eta_1-m}{\eta_1}T} \right)^2.
\end{equation}
Note that, when $m=0$, this reproduces the well-known result 
$p_{\circ}^{\max}=\eta_2(1-S)=\eta_2(1-|\braket{\phi_1}{\phi_2}|^2)$ 
for unambiguous discrimination in the case of $\eta_1 \le \eta_2 S$.  

We assumed that $\eta_1 \le \eta_2$. 
For the case of $\eta_1 \ge \eta_2$, it is clear that there is also a similar single-state 
domain, where $E_2$ is zero and state $\rho_2$ is omitted.

\section{Weak and strong error-margin conditions \label{sec:strong}}
Until this point, we considered the discrimination problem with an error margin 
imposed on the average probability of error $p_{\times}$. We can consider a different way 
of imposing an error margin. 
Suppose the measurement outcome is $\mu=1$. The probability of error in this case 
is the conditional probability $P_{\rho_2|E_1}$. 
In this section, we consider a discrimination problem with the conditions 
that the two conditional error probabilities must not exceed a certain error margin $m$.  
\begin{subequations}
   \label{eq:strong_margin}
\begin{align}
    P_{\rho_2|E_1} & \le  m, \\
    P_{\rho_1|E_2} & \le  m.                 
\end{align}
\end{subequations}
These conditions are stronger than the error-margin condition,  
Eq.~(\ref{eq:primal_weak_margin}), considered in preceding sections 
in the sense that Eq.~(\ref{eq:primal_weak_margin}) follows from 
Eqs.~(\ref{eq:strong_margin}). 
\begin{align*}
  p_{\times} &= P_{\rho_2,E_1}+P_{\rho_1,E_2} 
   = P_{\rho_2|E_1}P_{E_1} + P_{\rho_1|E_2}P_{E_2}  \\ 
   & \le  m ( P_{E_1}+P_{E_2} ) \le m. 
\end{align*}
We call the conditions given by Eqs.~(\ref{eq:strong_margin}) and 
Eq.~(\ref{eq:primal_weak_margin}) strong and weak error-margin conditions, respectively. 

For equal occurrence probabilities, optimal solutions have already been 
obtained for both the weak and strong error-margin conditions \cite{Hayashi08}. 
In the following, we will establish a relation between optimal solutions of 
the two error-margin conditions for general occurrence probabilities. 

In order to distinguish the two schemes, ``strong" and ``weak", we use superscripts 
$\S$ and $\W$, respectively. Let us start with the optimal POVM $E_\mu^\S(m^\S)$ with strong 
error-margin $m^\S$. Suppose we calculate average error probability by using 
$E_\mu^\S(m^\S)$, which we denote by $p_{\times}^\S(m^\S)$. 
Using conditional error probabilities, we observe  
\begin{align*}
  p_{\times}^\S(m^\S) &= P_{\rho_2,E_1}^\S(m^\S)+ P_{\rho_1,E_2}^\S(m^\S)\\ 
                      &= P_{\rho_2|E_1}^\S(m^\S)P_{E_1}^\S(m^\S) 
                         +P_{\rho_1|E_2}^\S(m^\S)P_{E_2}^\S(m^\S)\\
                      &\le  m^\S \left(  P_{E_1}^\S(m^\S) + P_{E_2}^\S(m^\S) \right) \\
                      &= m^\S \left( p_{\max}^\S(m^\S) + p_{\times}^\S(m^\S) \right), 
\end{align*}
from which it follows that 
\begin{equation*}
  p_{\times}^\S(m^\S) \le \frac{m^\S}{1-m^\S}p_{\max}^\S(m^\S).
\end{equation*}
This implies that the optimal POVM $E_\mu^\S(m^\S)$ with strong 
error-margin $m^\S$ satisfies the weak error-margin condition with 
$m^\W = \frac{m^\S}{1-m^\S}p_{\max}^\S(m^\S)$. 
Consequently, we obtain an inequality for two optimal success probabilities $p_{\max}^\S$ 
and $p_{\max}^\W$. 
\begin{equation}
  p_{\max}^\S(m^\S) \le p_{\max}^\W \left( \frac{m^\S}{1-m^\S}p_{\max}^\S(m^\S) \right). 
      \label{eq:inequality_I}
\end{equation}

Note that the relation $p_{\max}^\S(m) \le p_{\max}^\W(m)$ holds for a common value of 
error margin $m$, because the strong error-margin conditions are stronger than the weak 
error-margin condition. Here, however, inequality (\ref{eq:inequality_I}) involves 
error margins of different values, and it will be shown that equality actually holds 
in this inequality.  

We can derive another inequality for the two optimal success probabilities. 
Let us take the optimal POVM $E_{\mu}^\W(m^\W)$ satisfying a weak error margin $m^\W$. 
Remember that the two conditional probabilities of error are equal in the 
minimum-error and intermediate domains; $P_{\rho_2|E_1}^\W = P_{\rho_1|E_2}^\W$. 
In the single-state domain, one of the two conditional error probabilities is not defined. 
However, the following relations still hold with a constant $\kappa$:
\begin{align*}
      P_{\rho_2,E_1}^\W(m^\W) &= \kappa P_{E_1}^\W(m^\W), \\  
      P_{\rho_1,E_2}^\W(m^\W) &= \kappa P_{E_2}^\W(m^\W). 
\end{align*}
Adding these two expressions, we obtain 
\begin{align*}
  p_{\times}^\W(m^\W) &= \kappa \left( P_{E_1}^\W(m^\W)+P_{E_2}^\W(m^\W) \right) \\ 
                      &= \kappa \left( p_{\max}^\W(m^\W)+p_{\times}^\W(m^\W) \right),
\end{align*} 
from which it follows that 
\begin{align*}
   \kappa &= \frac{p_{\times}^\W(m^\W)}
                 {p_{\max}^\W(m^\W)+p_{\times}^\W(m^\W)} \\
          & \le  \frac{m^\W}{ p_{\max}^\W(m^\W) + m^\W}.  
\end{align*}
This shows that conditional error probabilities in the weak error-margin scheme 
satisfy the strong error margin conditions with $m^\S = \frac{m^\W}{ p_{\max}^\W(m^\W) + m^\W}$. 
We, therefore, obtain another inequality given by
\begin{equation}
  p_{\max}^\W(m^\W) \le p_{\max}^\S \left( 
                             \frac{m^\W}{p_{\max}^\W(m^\W)+m^\W} \right). 
             \label{eq:inequality_II}
\end{equation} 

Actually equality holds in inequalities (\ref{eq:inequality_I}) and (\ref{eq:inequality_II}). 
This can be seen by their repeated uses as follows: 
\begin{align*}
 & p_{\max}^\S(m^\S) \\
 & \le p_{\max}^\W \left( \frac{m^\S}{1-m^\S}p_{\max}^\S(m^\S) \right) \\
 & \le p_{\max}^\S \left(
        \frac{ \frac{m^\S}{1-m^\S} p_{\max}^\S(m^\S) }
             { p_{\max}^\W \left( \frac{m^\S}{1-m^\S}p_{\max}^\S(m^\S) \right) 
              + \frac{m^\S}{1-m^\S}p_{\max}^\S(m^\S) }
                     \right)        \\
 & \le p_{\max}^\S \left(
        \frac{ \frac{m^\S}{1-m^\S} p_{\max}^\S(m^\S) }
             { p_{\max}^\S(m^\S) + \frac{m^\S}{1-m^\S}p_{\max}^\S(m^\S) }
                     \right)        \\
 & = p_{\max}^\S(m^\S).  
\end{align*}
In the above derivation, we used the fact that the success probability is an increasing 
function of error margin. 

Thus, if two error margins $m^\S$ and $m^\W$ are related by 
\begin{equation}
  m^\S = \frac{m^\W}{p_{\max}^\W(m^\W)+m^\W},  \label{eq:mSmW}
\end{equation}
or equivalently by
\begin{equation}
   m^\W = \frac{m^\S}{1-m^\S}p_{\max}^\S(m^\S),
\end{equation}
the two optimal success probabilities are equal. 
\begin{equation}
   p_{\max}^\S(m^\S) = p_{\max}^\W(m^\W). \label{eq:PSPW}
\end{equation}
When one of the optimal success probabilities is known, the other can be 
determined by these equations. We note that the optimal POVMs are also related 
in the same way: $E_\mu^\S(m^\S)=E_\mu^\W(m^\W)$.

Using the above relation, we obtain the optimal success probability with the 
strong error-margin conditions to be    
\begin{align*}
 & p_{\max}^\S  = \\  
 &   \begin{cases} 
        \frac{1}{2}\left( 1 + \sqrt{1-4\eta_1\eta_2 S} \right) &
                            (m_c \le m \le 1), \\ 
        A_m \left( 1 - 2\sqrt{\eta_1\eta_2 S} \right)  &
                            (m_c' \le m \le m_c), \\ 
        \frac{\eta_1\eta_2(1-m)(1-S)}
             {m\eta_2+(1-m)\eta_1-2\sqrt{m(1-m)\eta_1\eta_2S}} &  
                            (0 \le m \le m_c'), \\
     \end{cases}
\end{align*}
where $A_m$ is given by
\begin{equation*}
    A_m = \frac{1-m}{(1-2m)^2} \left( 1+2\sqrt{m(1-m)} \right).
\end{equation*}
We assumed $\eta_1 \le \eta_2$, and $m_c$ and $m_c'$ are defined by      
\begin{align*}
 &   m_c  \equiv  \frac{1}{2}\left( 1 - \sqrt{1-4\eta_1\eta_2 S} \right),  \\
 &   m_c' \equiv  
            \begin{cases} 
               \displaystyle
               \frac{(\eta_1-\sqrt{\eta_1\eta_2S})^2}
                    {(\eta_2-\sqrt{\eta_1\eta_2S})^2 + (\eta_1-\sqrt{\eta_1\eta_2S})^2 } \\ 
               \hfill (\eta_1 \le \eta_2 S), \\
               0 \hspace{2ex} (\eta_1 \ge \eta_2 S).
            \end{cases} 
\end{align*}

\section{Upper bound for mixed state discrimination with error margin \label{sec:mixed}}
Let us consider that two states to be discriminated, $\rho_1$ and $\rho_2$, are mixed. 
The maximum success probability is known for minimum-error discrimination ($m=1$). 
For unambiguous discrimination ($m=0$) of general two mixed states, however, 
no analytic result for the maximum success probability is known. 
In Ref. \cite{Rudolph03}, Rudolph {\it et al}. presented a simple upper bound 
for the success probability, 
\begin{align}
 & p_{\max}(\rho_1,\rho_2)  \nonumber \\ 
 & \le \begin{cases} 
          1 - 2\sqrt{\eta_1\eta_2 }F(\rho_1,\rho_2) &
                       (\eta_1 \ge \eta_2 F(\rho_1,\rho_2)^2), \\ 
          \eta_2 \left( 1 - F(\rho_1,\rho_2)^2 \right) & 
                       (\eta_1 \le \eta_2 F(\rho_1,\rho_2)^2). \\
       \end{cases} 
            \label{eq:known_upper_bound}
\end{align} 
where $F(\rho_1,\rho_2)=\tr(\sqrt{\rho_1}\rho_2\sqrt{\rho_1})^{1/2}$ 
is the fidelity of states $\rho_1$ and $\rho_2$. 
Later, the conditions for the two mixed states to reach the upper bound 
were analyzed and a new series of upper bounds was also found 
(see e.g., Refs. \cite{Feng04, Herzog05, Raynal05, Zhou07}).   
For general error margin, a closed form of the maximum success probability is 
also hard to obtain as in unambiguous discrimination.  
However, it is likely that there exists an upper bound similar to 
Eq.~(\ref{eq:known_upper_bound}), since it is expressed in terms of the fidelity of 
the two states and their occurrence probabilities. 
In the following, we will show that the method of Rudolph {\it et al.}  
can be applied to the case of general margin and an upper bound for success probability can 
easily be obtained by using the results of pure-state discrimination. 

Suppose states to be discriminated are prepared in system $Q$, and 
purify the states by introducing another system $R$ \cite{Nielsen_text_book}. 
\begin{align}
   \rho_1^Q &= \trm_R  \ket{\Psi_1^{QR}}\bra{\Psi_1^{QR}},\\
   \rho_2^Q &= \trm_R  \ket{\Psi_2^{QR}}\bra{\Psi_2^{QR}}. 
\end{align}
We assume that pure states $\ket{\Psi_1^{QR}}$ and $\ket{\Psi_2^{QR}}$ are 
chosen so that
\begin{equation*}
    |\braket{\Psi_1^{QR}}{\Psi_2^{QR}}| = F(\rho_1,\rho_2),
\end{equation*}
which is always possible by Uhlmann's theorem \cite{Uhlmann76}.   

Consider a hypothetical discrimination problem between pure states 
$\ket{\Psi_1^{QR}}$ and $\ket{\Psi_2^{QR}}$ with occurrence probability 
$\eta_1$ and $\eta_2$, respectively. We take the weak error-margin condition. 
The task is to maximize the success probability 
\begin{align}
  p_{\circ} \equiv \eta_1 & \trm_{QR}E_1^{QR}\ket{\Psi_1^{QR}}\bra{\Psi_1^{QR}} \nonumber \\ 
                   &+ \eta_2 \trm_{QR}E_2^{QR}\ket{\Psi_2^{QR}}\bra{\Psi_2^{QR}}, 
            \label{eq:p_circ_QR}
\end{align}
under the condition that the average probability of error  
\begin{align}
 p_{\times} \equiv \eta_1 & \trm_{QR}E_2^{QR}\ket{\Psi_1^{QR}}\bra{\Psi_1^{QR}} \nonumber \\ 
                  &+ \eta_2 \trm_{QR}E_1^{QR}\ket{\Psi_2^{QR}}\bra{\Psi_2^{QR}}, 
            \label{eq:p_times_QR}
\end{align}
must not exceed error margin $m$. 
The maximum success probability for two pure states $\ket{\phi_1}$ and $\ket{\phi_2}$ 
is a function of $|\braket{\phi_1}{\phi_2}|$ and independent of the dimension.  
We denote it by $p_{\max}^{\rm pure}(|\braket{\phi_1}{\phi_2}|)$.  
The maximum success probability for the hypothetical discrimination problem is then  
given by $p_{\max}^{\rm pure}(F(\rho_1,\rho_2))$. 

Let us impose an extra constraint on POVM $E_{\mu}^{QR}$ in this discrimination 
problem: 
\begin{equation}
     E_{\mu}^{QR} = E_{\mu}^{Q} \otimes \mbold{1}^{R},\ \mu=1,2,3.
\end{equation}
By this additional condition, the success probability (\ref{eq:p_circ_QR}) and the 
average error probability (\ref{eq:p_times_QR}) are reduced to 
\begin{align*}
  p_{\circ} &=  \eta_1 \trm_Q E_1^Q \rho_1^Q + \eta_2 \trm_Q E_2^Q \rho_2^Q, \\
  p_{\times} &= \eta_1 \trm_Q E_2^Q \rho_1^Q + \eta_2 \trm_Q E_1^Q \rho_2^Q,
\end{align*}
and the problem becomes equivalent to discrimination between 
the two mixed states $\rho_1$ and $\rho_2$ with occurrence 
probabilities $\eta_1$ and $\eta_2$. 
It is clear that any extra condition on POVM never increases the maximum success probability. 
Thus, we conclude that the success probability for two mixed states is upper-bounded 
by the maximum pure-state success probability with $|\braket{\phi_1}{\phi_2}|$ 
replaced by the fidelity of the two mixed states. 
\begin{equation*}
   p_{\max}(\rho_1,\rho_2) \le p_{\max}^{{\rm pure}}(F(\rho_1,\rho_2)).
\end{equation*}

Using the results of pure-state discrimination given in Eq.~(\ref{eq:weak_maximum_p_circ}), 
we obtain  
\begin{align*}
 & p_{\max}(\rho_1,\rho_2)    \\
 &\le \begin{cases} 
        \left( \sqrt{m} + \sqrt{ 1 - 2\sqrt{\eta_1\eta_2 }F(\rho_1,\rho_2) } \right)^2  \\
             \hspace{25ex}   (m_c' \le m \le m_c), \\ 
        \eta_2 \left( \sqrt{\frac{m}{\eta_1}}F(\rho_1,\rho_2) 
               + \sqrt{\frac{\eta_1-m}{\eta_1}(1-F(\rho_1,\rho_2)^2)} \right)^2 \\ 
             \hspace{25ex}   (0 \le m \le m_c'), \\
      \end{cases}
\end{align*}
where $m_c$ and $m_c'$ are given by 
\begin{align*}
    m_c &= \frac{1}{2}\left( 1 - \sqrt{1-4\eta_1\eta_2F(\rho_1,\rho_2)^2} \right),  \\
    m_c'&=  \begin{cases} 
               \displaystyle
               \frac{\left(\eta_1-\sqrt{\eta_1\eta_2}F(\rho_1,\rho_2)\right)^2}
                    {1-2\sqrt{\eta_1\eta_2}F(\rho_1,\rho_2)} &
                            (\eta_1 \le \eta_2 F(\rho_1,\rho_2)^2), \\
               0 & (\eta_1 \ge \eta_2 F(\rho_1,\rho_2)^2).
            \end{cases}
\end{align*} 
We assumed $\eta_1 \le \eta_2$ as in the pure-state case. 

For unambiguous discrimination ($m=0$), the upper bound is reduced to the one 
given in Eq.~(\ref{eq:known_upper_bound}).
The maximum success probability of minimum-error discrimination is 
known and given by   
\begin{equation*}
   p_{\max}(\rho_1,\rho_2) = 
          \frac{1}{2} \left( 1 + \tr{\left| \eta_1\rho_1-\eta_2\rho_2 \right| \right)}, 
\end{equation*} 
which must not exceed our upper bound. This observation leads to an inequality 
\begin{equation*}
  \tr{\left| \eta_1\rho_1-\eta_2\rho_2 \right|}
   \le \sqrt{1-4\eta_1\eta_2F(\rho_1,\rho_2)^2}, 
\end{equation*}
which is a generalization of the well-known inequality concerning the trace distance 
and the fidelity \cite{Nielsen_text_book}, 
\begin{equation*}
 \frac{1}{2} \tr{|\rho_1-\rho_2|} \le \sqrt{1-F(\rho_1,\rho_2)^2}. 
\end{equation*}
  
\section{Concluding remarks}
In this paper, we considered a state discrimination problem which interpolates 
minimum-error and unambiguous discriminations by introducing a margin 
for the probability of error. In the case of two pure states with general occurrence 
probabilities, we obtained the optimal success probability in a fully analytic form. 

Our final remark is about the possibility of optimal local discrimination 
between two multipartite pure states. Suppose two pure states are multipartite 
and generally entangled. An interesting question is whether the parties sharing 
the states can achieve the globally optimal success probability by local operations 
and classical communication (LOCC). 
It is known that two pure states can be optimally discriminated by LOCC in both 
the minimum-error \cite{Walgate00,Virmani01} and unambiguous \cite{Chen02,Ji05} 
discrimination schemes. For general error margin, we showed that this is also true 
when the occurrence probabilities are equal \cite{Hayashi08}.  
To show this, we proved the following general theorem \cite{Hayashi08}:  

\smallskip\noindent{\bf Theorem}: {\it 
Let $V$ be a two-dimensional subspace of a multipartite tensor-product space $H$, and $P$ be 
the projector onto the subspace $V$. Then, for any three-element POVM 
$\{E_1,E_2,E_3\}$ of $V$ with every element being of rank 0 or 1, there exists a one-way LOCC POVM 
$\{E_1^{{\rm L}},E_2^{{\rm L}},E_3^{{\rm L}}\}$ of $H$ such that  
$
  E_\mu = P E_\mu^{{\rm L}} P \ (\mu=1,2,3). 
$
}\smallskip

This implies that a POVM satisfying the conditions of Theorem can be implemented 
by a one-way LOCC protocol as far as measurement for states in subspace $V$ is  
concerned. As we have seen in Sec.~\ref{sec:problem}, for general occurrence 
probabilities, the optimal POVM elements are also of rank 1 at most.  
Thus, for any error margin and any occurrence probabilities, two multipartite pure states 
can be optimally discriminated by LOCC.


\begin{acknowledgments}
A.H. would like to thank Masahito~Hayashi for fruitful discussions 
and his valuable suggestion on the relation between the weak and strong 
error-margin conditions.  
\end{acknowledgments}


\end{document}